\documentclass[conference,letterpaper]{IEEEtran}
\usepackage{amsfonts}

\usepackage[margin=0.63in]{geometry}
\usepackage{enumitem}
\usepackage{algorithm,algorithmic}
\usepackage{algorithmic}
\usepackage{graphicx}
\usepackage{subcaption}
\usepackage{caption}
\DeclareMathAlphabet\mathbfcal{OMS}{cmsy}{b}{n}

\newcommand{\A}{\mathcal{A}}

\newcommand{\Se}{\mathcal{S}}

\newcommand{\nix}[1]{}

\usepackage[utf8]{inputenc} 
\usepackage[T1]{fontenc}
\usepackage{url}
\usepackage{ifthen}
\usepackage[cmex10]{amsmath} %

\usepackage{algorithm}
\usepackage{tikz}
\usepackage{pgfplots}
 \usepackage{url}

\usepackage[hidelinks,colorlinks=true,urlcolor= blue,linkcolor = blue,citecolor = red]{hyperref}

\usepackage{color}

\usepackage{amssymb,paralist}

\newtheorem{theorem}{Theorem}

\newtheorem{definition} {Definition}

\newtheorem{remark}{Remark}

\makeatletter
\newcommand*{\rom}[1]{\expandafter\@slowromancap\romannumeral #1@}
\makeatother

\usepackage[numbers,sort,comma,round]{natbib}

\begin{document}

\title{Capacity Achieving Uncoded PIR Protocol based on Combinatorial Designs}

\author{%
  \IEEEauthorblockN{Mohit Shrivastava and Pradeep Sarvepalli}
  \IEEEauthorblockA{Department of Electrical Engineering\\
                    Indian Institute of Technology Madras\\
                    Chennai, India 600 036\\
                    }

}

\maketitle

\begin{abstract}

In this paper we study the problem of private information retrieval where a user seeks to retrieve one of the $F$ files from a cluster of $N$ non-colluding servers without revealing the identity of the requested file. In our setting the servers are storage constrained in that they can only store a fraction $\mu=t/N$ of each file. Furthermore, we assume that the files are stored in an uncoded fashion. The rate of a PIR protocol is defined as the ratio of the file size and the total number of bits downloaded. The maximum achievable rate is referred to as capacity. It was previously shown that there are capacity achieving PIR protocols when the file size is $N^F$ and complete files were stored on all the servers. These results were further extended for the case when servers store only a fraction of each file. However, the subpacketization $v$ of the files  required is exponential in the number of servers $N$.  We propose a novel uncoded PIR protocol based on combinatorial designs that are also capacity achieving when the file size is $v \times t^F$. Our protocol has linear subpacketization in the number of servers in contrast to previous work in storage constrained uncoded PIR schemes. In the proposed PIR protocol, the given system is projected to multiple instances of reduced systems with replicated servers having full storage capacity. The subfiles stored in these various instances are separately retrieved and lifted to solve the PIR problem for the original system. 
\end{abstract}

\section{Introduction}

The notion of Private Information Retrieval (PIR) was first introduced by Chor \textit{et al.} in \cite{chor95}, and since then, the field has immensely grown   \cite{henry11,shah14,sun17,sun17b,banawan18,kumar19,attia20}. 
PIR systems allow a user to query a cluster of servers and retrieve the desired file without revealing any information about the desired file index to any of the individual servers.
The elemental setting of PIR considers the servers to be non-colluding i.e. the servers cannot cooperate with each other to recover information about the query.  
PIR becomes important sometimes to protect users from surveillance, monitoring and profiling.

In such a scenario it becomes important that the querying of servers by the user does not leak any information about the user's requirements. 

One can achieve privacy by downloading all the files. 
Clearly, this is an inefficient method to achieve privacy. 
A chief problem of PIR protocols is to achieve privacy efficiently. 
Efficiency is typically measured in terms of the rate of the protocol. 
The ratio of size of retrieved file to the amount to data downloaded is referred to as the PIR rate.  A higher PIR rate signifies higher efficiency, and the highest achievable PIR rate is referred to as the PIR capacity. The reciprocal of PIR rate is referred to as download cost per bit and is a measure of number of bits that need to be downloaded to retrieve one bit of the desired file. The optimal download cost per bit is the lowest achievable download cost per bit.

\noindent
{\em Previous work.}
PIR protocols can store the files in a coded form or an uncoded form. 
In the recent years there has been a growing interest in uncoded PIR due to the simplicity of encoding and decoding the files and also because they can offer competitive performances compared to coded protocols \cite{sun17,attia20,zhang19}.
In this paper we are interested in uncoded PIR protocols. 
We consider a system of $N$ non-colluding servers, each storing $\frac{t}{N}$ fraction of each of the $F$ independent files. 

The capacity of uncoded PIR from $N$ non-colluding, replicated databases, each storing all $F$ files completely was characterized by Sun and Jafar in \cite{sun17}.
They also provided a  capacity achieving protocol for file sizes $L=N^{F}$.
 Attia \textit{et al.} \cite{attia18} showed that, for any uncoded PIR scheme from storage constrained servers, the achievable rate $R$ is bounded above by $\big( 1+\frac{1}{t}+\dots+\frac{1}{t^{F-1}}\big)^{-1}$, where $F$ is the number of files in the system and every server stores fraction $\frac{t}{N}$ of each file, $N$ being the number of servers. The capacity achieving scheme they proposed requires a subpacketization of the file into $\binom{N}{t}$ pieces, which is exponential in the number of servers $N$. 
Subsequently,   
Zhang, \textit{et al.} \cite{zhang19}, improved upon the subpacketization and achieved capacity using a $t$ times smaller subpacketization, which is still exponential in the number of servers $N$.

\noindent
{\em Contributions. } 
We propose a novel PIR protocol for uncoded systems storing a fraction of each file.
While the storage capacity is identical for the servers, they need not store the same content. 
The proposed PIR protocol has two main components: a storage scheme and a retrieval scheme. 
The storage scheme is based on combinatorial designs, more specifically tactical configurations, 
while the retrieval scheme works by projecting the given system to multiple instances of smaller systems with replicated servers having full storage capacity.
It uses the retrieval scheme of \cite{sun17} for these smaller instances. 
Our scheme is flexible and can easily accommodate a range of system parameters. 

The proposed protocol also achieves linear subpacketization in contrast to the exponential subpacketization required in previous works \cite{attia18,zhang19}. 

The rest of this paper is organized as follows. In Section~\ref{sec:bg}, we review the necessary background.
Then in Section~\ref{sec:prop-PIR}, we present our central result, namely a construction of PIR protocol. 
We present the performance of our protocol in Section~\ref{sec:performance}. Finally we conclude and discuss further directions in Section~\ref{sec:conc}. %

\section{System Model}\label{sec:bg}
In this section, we review the system model for PIR systems and some relevant work. 
We also provide a brief review of some necessary ideas from combinatorial design theory. 
For a positive integer $n$, we 
define $[n]\triangleq \{1, 2, \ldots, n \}$.

\subsection{Setup for PIR}
We consider a system of $N$ non-colluding servers storing $F$ independent files. 
The servers are denoted as $\Se_n$ where $n\in [N]$.
Each file is assumed to be of size $L$ bits. 
The files are denoted as $W_i$, where $i\in [F]$.

As the files are independent we have $H(W_1,W_2,\dots,W_F) =\sum_{i=1}^n H(W_i) = F L$,
where $H(\cdot )$ denotes the entropy function. 

We   assume that the servers have identical storage capacity and every server stores 
a fixed fraction $\mu \in [\frac{1}{N},1]$ of every file.
This fraction $\mu$ is called the normalized storage capacity of the server.
Therefore each server stores $\mu L$ bits of each file and a total of $\mu L F$ bits.
For simplicity we consider the case where $\mu$ is an integral multiple of $1/N$.
PIR protocols can be extended to other values of $\mu$ by memory sharing, see for instance \cite{attia20}. 

\begin{figure}[H]
\centering
\includegraphics[scale=0.75]{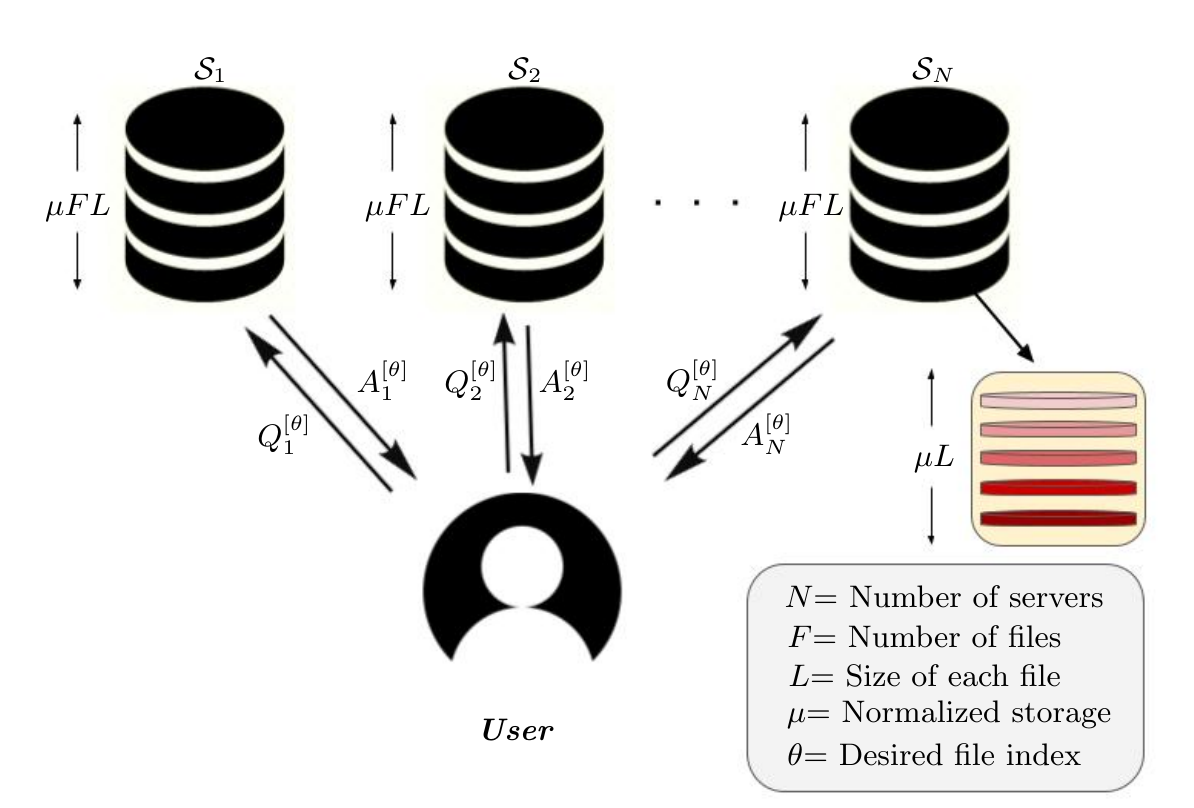}
\captionsetup{justification=justified}
\caption{ A $\mu$-$(F, N )$ system with $N$ servers storing $F$ independent files.
The user generates an index $\theta \in [F]$ and a query $Q_n^{[\theta]}$ to each of the servers.
The server responds with an answer denoted as $A_n^{[\theta]}$. 
The user recovers the requested file from all the answers. 
} \label{fig:pir-blockdiagram}
\end{figure}

We denote the content of server $S_n$ by $Z_{n}$, for any $n \in [N]$. 
A  system of  $N$ non-colluding servers, storing $F$ files and having a normalized storage of $\mu$ is referred to as a $\mu$-$(F, N)$ system,
see  Fig.~\ref{fig:pir-blockdiagram} for an illustration.

The PIR problem is described as follows. A user privately generates 
an index $\theta \in [F]$ and wishes to retrieve file $W_\theta$, without revealing any information about $\theta$ to any of the individual servers. 
This index $\theta$ is generated independent of the file content or the server contents. 
To retrieve $W_\theta$, the user generates $N$ queries $Q_1^{[\theta]},Q_2^{[\theta]},\dots,Q_N^{[\theta]}$, where $Q_n^{[\theta]}$ is sent to server $\Se_n, \;n\in [N]$. 
Upon receiving these queries, each server $\Se_n$ responds with an answer $A_n^{[\theta]}$, which is a deterministic function of the query and the contents stored on the server, i.e.,
\begin{equation}
    H(A_n^{[\theta]}| Q_n^{[\theta]},Z_{n})=0, \text{ for all } n\in [N] \text{ and } \theta \in [F]. \label{eq:pir-constraint-0}
\end{equation}
In this work, we assume that  the user requests a linear combination of bits stored on the server as the query.
The server responds by returning the value of the requested linear combination. 

A PIR protocol must satisfy the following constraints:
\begin{subequations}
\begin{eqnarray}
    H(W_\theta | A_1^{[\theta]},\dots,A_N^{[\theta]},Q_1^{[\theta]},\dots,Q_N^{[\theta]})=0,\text{ for all }\theta \in [F].    \label{eq:pir-recovery}\\
    I(\theta;Q_n^{[\theta]},A_n^{[\theta]},Z_n)=0, \;\;\forall n\in [N],\; \text{ for all } \theta \in [F] \label{eq:privacy-condition}
\end{eqnarray}
\end{subequations}
The first condition \eqref{eq:pir-recovery} requires that 
from the answers obtained form all the servers, the user must be able to correctly retrieve the desired file $W_\theta$.

The second condition \eqref{eq:privacy-condition} ensures privacy of request i.e.
the user does not reveal any information about the desired file index to any of the individual servers.

For any PIR protocol, the total size of all queries sent to the servers constitutes the upload cost and the amount of downloaded data constitutes the download cost. 
The total cost of PIR is the sum of upload cost and download cost, with the later usually being the dominant contributor. 
Thus, it is common to ignore the upload cost while quantifying the efficiency of a PIR protocol. 
The following results are known regarding the capacity of PIR protocols. 

\begin{theorem}[Attia {\em et al.} \cite{attia20}]{\label{th:cap-mu}}
    For the $\mu$-$(F, N)$  system defined above, the optimal download cost per bit  $D^\ast(\mu)$ (or the inverse of PIR capacity) is given by the lower convex hull of the following $(\mu,D^{*}(\mu))$ pairs, for $t=1,2,\dots,N$: 
    \begin{eqnarray}
    \bigg(\; \mu=\frac{t}{N},\; D^{*}(\mu)=\sum_{f=0}^{F-1} \frac{1}{t^f}\; \bigg).\label{eq:pir-capacity}
    \end{eqnarray}
\end{theorem}

For $\mu=1$, the optimal cost $D^{*}=\sum_{f=0}^{F-1}\frac{1}{N^f}$, which means the maximum achievable rate (capacity) $R^{*}=\big(1+\frac{1}{N}+\dots+\frac{1}{N^{F-1}}\big)^{-1}$,  originally shown by Sun and Jafar in \cite[Theorem~1]{sun17}).
For comparison, the optimal download cost of PIR for a coded  $\mu$-$(F, N)$ system using MDS codes was characterized by Tajeddine \textit{et al.} in \cite{tajeddine18}.
Using $(N, K)$ MDS codes they show that
\begin{eqnarray}
D^{*}(\mu_{MDS})=1+\frac{K}{N}+\frac{K^2}{N^2}+\dots +\frac{K^{F-1}}{N^{F-1}}\label{eq:pir-cap-mds}
\end{eqnarray}
 where $\mu_{MDS}=1/K$.

\subsection{Designs}

The storage scheme of the PIR protocol we propose in  Section~\ref{sec:prop-PIR} is based on combinatorial designs. 
So, we review some basic definitions and properties of combinatorial designs. 
%
We refer the reader to \cite{stinson07} for more details.

\begin{definition}[Design] 
A design is a pair $(X,\A)$ such that the following are satisfied:
\begin{compactenum}[D1.]
    \item $X$ is a set of elements called points.
    \item $\A$ is a collection of non-empty subsets of $X$ called blocks.
\end{compactenum}
\end{definition}

A design can also be described in terms of a matrix called the incidence matrix. 
We define this next. 
\begin{definition}[Incidence Matrix]
Let $(X,\A)$ be a design where $X=\{ x_1,x_2,\dots,x_{v}\}$ and $\A=\{A_1,A_2,\dots,A_{b}\}$. The incidence matrix of $(X,\A)$ is the $v \times b$ binary matrix $M=(M_{ij})$ defined by the rule 
\begin{equation}
    M_{ij} =
\left\{
	\begin{array}{ll}
		1  & \mbox{if } x_i \in A_j \\
		0  & \mbox{if } x_i \notin A_j
	\end{array}
\right.   \label{eq:incidence-matrix}
\end{equation}
\end{definition}

Of particular interest to us are a class of designs called tactical configurations.

\begin{definition}[Tactical configuration] 
A ($v,k,b,r$)-configuration, also known as tactical configuration is a design with $v$ points and $b$ blocks, each containing $k$ points. Also, each point is contained in the same number of blocks $r$, called the repetition number. 
\end{definition}

In \cite{bill}, it was shown that the necessary and sufficient condition for existence of a ($v,k,b,r$)-configuration is 
\begin{equation}
    b k=v r, \text{ where } v,k,b,r \in \mathbb{Z}^{+} \label{existence config}
\end{equation}

The incidence matrix $M$ of a ($v,k,b,r$)-configuration has some useful structure:
\begin{compactenum}[(i)]
    \item Each row of $M$ has exactly $r$ ones.
    \item Each column of $M$ has exactly $k$ ones.
\end{compactenum}


\section{PIR Protocol from Designs} \label{sec:prop-PIR}

In this section we explain the storage and retrieval schemes for the proposed protocol with examples. We prove that the protocol is both private and correct. Also, we characterize the rate and subpacketization for the proposed protocol.

\subsection{Proposed PIR protocol}

Given a system with $N$ servers, each with a normalized storage $\mu=t/N$ and a set of $F$ independent files, the storage scheme is given as follows:
\begin{algorithm}[H]
\caption{Storage scheme for proposed PIR protocol\\\textbf{Input}: $N,t$\\\textbf{Output}: $Z_n$, $\text{ for all } n\in [N]$ }
{\label{alg:storage}}
  \begin{algorithmic}[1]
    \STATE Construct a ($v,k,b,r$)-configuration with $b=N$ and $r=t$. Denote the  ${v \times b}$ incidence matrix of this configuration by $M$ whose $(i, j)$th entry is given by  $M_{ij}$.
    \STATE  Divide each file $W_i$, for $i \in [F]$, into $v$ subfiles. Denote them as
    $W_{ij}$ where $j \in [v]$. 
    Each subfile  $W_{ij} $ is of size $L/v$ bits. 
    \STATE For all $i\in [F]$ and $j\in [v]$ store the subfile $W_{ij}$ on server $\Se_n$ if $M_{jn}=1$.
  \end{algorithmic}
\end{algorithm}

Note that each server stores $k$ out of the $v$ subfiles for each file (since each column of $M$ has exactly $k$ ones). 
Thus, the normalized storage of each server is $\mu=k/v=r/b=t/N$.

     Let $\Se(W_{i j})$ be the subset of servers that store the subfile $W_{i j}$ for a fixed $i, j$ where $i\in[F], j\in [v]$.
Each subfile $W_{ij}$ is present on exactly $t$ servers (since each row of $M$ has exactly $r=t$ ones). 
In other words, $|\Se(W_{i j})|=t$.

For any fixed $j\in [v]$, we denote the part of the  $Z_n$ containing the $j^{th}$  subfiles $W_{ij}$  as $Z_n^{[j]}$.

\begin{algorithm}[H]
\caption{Retrieval scheme for proposed PIR protocol\\\textbf{Input}: $\theta$\\\textbf{Output}: $W_\theta$ }
{\label{alg:recovery}}
  \begin{algorithmic}[1]
    \STATE For all $j\in [v]$, consider the system $\Se(W_{\theta j})$, which is a 1-($F,t$) system storing $W_{ij}$, $i\in [F]$. 
    \STATE Using the protocol in \cite{sun17}, generate queries $Q_n^{[\theta j]}$, for all $ \Se_n\in \Se(W_{\theta j})$.
    \STATE Using the answers $A_n^{[\theta j]}$ retrieve all bits of desired file $W_{\theta j}$  
   
  \end{algorithmic}
\end{algorithm}

\subsection{Correctness and privacy of the proposed protocol}

Suppose the user wants to retrieve the file $W_\theta$. 
To retrieve  file $W_\theta$, the user needs to retrieve all the subfiles $W_{\theta j}$ for $j \in[v]$.
We can break it down into $v$ instances of retrieval of the subfiles of $W_\theta$. 
The retrieval of each subfile $W_{\theta j}$ can be treated as a separate instance of a 
PIR retrieval problem.

\begin{theorem}[Designs to PIR protocols]\label{th:design2pir}
     Consider a $\mu$-$(F, b)$ system generated from a $(v, k, b, r)$-tactical configuration. 
     Then with respect to the subfiles $W_{ij}$, where $i\in [F]$, the servers $\Se(W_{\theta j})$ form a 
     1-$(F, r)$  system. 
     For file size $L=v \times r^F$, the proposed protocol 
     consisting of Algorithms~\ref{alg:storage}~and~\ref{alg:recovery} is correct, private and  achieves capacity.
\end{theorem}
\begin{IEEEproof}
In the $\mu$-$(F, b)$ system the subfile $W_{\theta j}$ is present on $r$ servers in the set
$\Se(W_{\theta j})$.
Furthermore, each of the servers in $\Se(W_{\theta j})$ contains all the subfiles 
$W_{ij}$ for all $i\in [F]$. 
Therefore, with respect to the subfiles $W_{ij}$, $i\in [F]$, the servers  $\Se(W_{\theta j})$ form a 1-$(F, r)$  system.

For recovering $W_{\theta j}$, we only restrict our attention to the 1-$(F, r)$  system obtained by restricting to the servers $\Se(W_{\theta j})$. 
Subfile $W_{\theta j}$ can be retrieved privately from the above system by using the protocol of \cite{sun17}.
Since $W_{\theta j}$ can be recovered for all $j\in [v]$ from the 
$1$-$(F,r)$ systems formed by $\Se(W_{\theta j})$, we are able to recover the file 
$W_\theta$.
Therefore, \eqref{eq:pir-recovery} is satisfied.

We note two properties of the proposed protocol that we need to prove the privacy constraint.
\begin{compactenum}[P1)]
    \item If we interchange all bits of file $W_1$ with $W_i$ where $i\in [F]$ in the
    queries to all the servers, we would retrieve file $W_i$ instead of $W_1$. 
    This is because the storage is symmetric with respect to all the files.   \item Any permutation of bits of $W_{ij},\;i\in[F],j\in[v]$ applied on  all the queries to all the servers does not affect the retrieval process, since permutation is an invertible operation. This is equivalent to choosing a different permutation in the protocol of \cite{sun17}.
\end{compactenum}

Next to show privacy of  recovery process, we introduce the following notation. 
Suppose to recover the subfile $W_{\theta j}$,
the protocol for $1$-$(F, r)$ system storing $W_{\theta j}$ generates queries $Q_{n}^{[\theta j]}$,  to send to server $\Se_n$, wherein $\Se_n \in \Se(W_{\theta j})$.  
Then denote the responses of $\Se_n$ as 
 $A_{n}^{[\theta j]}$, $\Se_n \in \Se(W_{\theta j})$. 
 For $\Se_n$ such that $\Se_n \notin \Se (W_{\theta j})$, we define  $Q_{n}^{[\theta j]}=\emptyset$.
We can combine all the queries sent by the user to recover $W_\theta$
as 
\begin{equation}
     Q_{n}^{[\theta]} = \bigcup_{j \in [v]}  Q_{n}^{[\theta j]} \label{eq:query-all}
\end{equation}
Likewise, we can combine all the responses 
from server $S_n$ as 
\begin{eqnarray}
A_n^{[\theta]} &= & \bigcup_{j\in [v]} A_n^{[\theta j]}
\end{eqnarray}

To show that the protocol is private, we need to show \eqref{eq:privacy-condition}.
Since the subfile $W_{\theta j}$ can be recovered privately, none of the servers can infer anything about $\theta$ given the queries, answers and stored content. 
Therefore we have 
\begin{align}
       I(\theta;Q_{n}^{[\theta j]},A_{n}^{[\theta j]},Z_{n}^{[j]})=0 \text{ for all } \Se_n \in \Se(W_{\theta j}) \label{privacy a1}
\end{align}
From this it follows that $I(\theta;Q_{n}^{[\theta j]})=0 \text{ for all } \Se_n \in \Se(W_{\theta j})$.
If the subfile $W_{\theta j}$ is not present on $\Se_n$, then $Q_n^{[\theta j]}$ is a null query
and once again we have $I(\theta;Q_{n}^{[\theta j]})=0 $.
\begin{align}
         I(\theta;Q_{n}^{[\theta j]})=0 \text{ for all } n\in [b] \label{privacy a2}
\end{align}
Now, for any server $\Se_n$, $n\in [b]$ 
\begin{align}
   I(\theta;Q_{n}^{[\theta]},A_{n}^{[\theta]},Z_{n}) & = I(\theta;Q_{n}^{[\theta]},Z_{n})  \label{privacy b1}
\end{align}
since $A_n^{[\theta]}$ is a deterministic function of $Q_n^{[\theta]}$ and $Z_n$.
Therefore, using \eqref{eq:query-all} we can write 
\begin{align}
   I(\theta; &Q_{n}^{[\theta]},  A_{n}^{[\theta]}, Z_{n})
   =  I(\theta;\underset{j \in [v]}{\cup}Q_{n}^{[\theta j]} ) 
   + I(\theta;Z_{n}|Q_{n}^{[\theta ]}) \nonumber\\
   &\overset{(a)}{=}  I(\theta;\underset{j \in [v]}{\cup}Q_{n}^{[\theta j]} ) + H(Z_n|Q_{n}^{[\theta]})
   - H(Z_n|\theta, Q_{n}^{[\theta]}) \nonumber \\
   &\overset{(b)}{=} I(\theta;\underset{j \in [v]}{\cup}Q_{n}^{[\theta j]} ) +H(Z_n) - H(Z_n) = I(\theta;\underset{j \in [v]}{\cup}Q_{n}^{[\theta j]} ) \nonumber
\end{align}
The last equality $(b)$  follows from the fact that  $\theta$ and $Q_n^{[\theta]}$ are user generated quantities (generated without any communication to the server) and hence cannot contain any information about the server contents $Z_n$.

It remains to show that $I(\theta;\underset{j \in [v]}{\cup}Q_{n}^{[\theta j]} ) =0$.


Without loss of generality, assume that $\theta=1$ and $n=1$ and set $q:=Q_1^{[1]}$.

We will show that, fixing $q$ to server $\Se_1$ the user can retrieve not only file $W_1$, but it can actually retrieve any file $W_i,\;i\in [F]$ by appropriately altering the queries to other servers. Also, we show that the probability of recovering any file $W_i,\;i\in [F]$, given $\Se_1$ receives query $q$, is the same. 
Showing these will establish that $I(\theta; Q_n^{[\theta]})=0,\;n\in [N]$. 
We need the following properties of the PIR protocol from \cite{sun17}, see Lemma~1 therein.
\begin{compactenum}[SJ1)]
    \item There are exactly $r^{F-1}$ bits for each subfile $W_{ij}$, involved in each query block. 
    In other words,  $r^{F-1}$ bits of $W_{ij}$, $\forall \;i \in [F]$ are involved in the $j^{th}$ query block to $\Se_n \in \Se(W_{ij})$.
    \item Any bit appears atmost once in the queries sent to a particular server, i.e. a bit involved in any query to $\Se_n$ doesnot appear in any other query to $\Se_n$.
    \item The query structure is symmetric with respect to any file. 
    In other words, the linear combinations are similar and only differ in the actual bits 
    forming the queries.

\end{compactenum}

To alter the queries to other servers so that we can recover a different file we proceed as follows. 

\begin{compactenum}[i)]
    \item  Interchange all bits of file $W_1$ with $W_i$ in the queries to all the servers. 
    This gives a (new) set of queries to be sent to the servers using which we can retrieve $W_i$. 
    \item Let $q_{new}$ be the query to $\Se_1$ in the new set of queries. 
    Next, we try to make all queries in the query sets $q$ and $q_{new}$ same.
    Recall that by SJ1) the same number of bits of $W_{ij}$ occur in any query block to a server and by SJ3) all query blocks have the same structure. 
    Therefore, by permuting the bits of $W_{ij}$ we can map  $q_{new}$ to $q$ uniquely
    with respect to the variables of $q$ and $q_{new}$. 
    This does not affect the recovery of the file $W_i$.
\end{compactenum}
This establishes that with respect to the queries of $\Se_1$, the user can modify the queries to other servers so that any file $i\in [F]$ can be recovered. 
Furthermore, the variables which are not part of $q$ and $q_{new}$ can also be permuted without affecting the recovery. 
Since the number of variables not involved in these queries are same for all files, 
from the point of view of the server there is equal uncertainty as to which file was requested by the user. 
These permutations exhaust all the possibilities of queries consistent with the PIR 
protocol  of \cite{sun17}.
Therefore, \eqref{eq:privacy-condition} is also satisfied and the proposed protocol is private. 



Now the size of the subfile is $r^F$, so the rate of the PIR protocol for the $1$-$(F, r)$ system is ${R}= (\sum_{f=0}^{F-1}r^{-f})^{-1}$.
Then to recover each of the subfiles $W_{\theta j}$ for a given index $\theta$ we require
to download $L/v{R}$ bits. 
For recovery of the entire file we need to download $L/{R}$ bits. 
Thus the rate of the proposed PIR protocol for the $\mu$-$(F, b)$ system is $L/(L/{R})= {R}$ which by Theorem~\ref{th:cap-mu} coincides with the capacity of the $\mu$-$(F, b)$ system when $L=v\times r^f$.
\end{IEEEproof}


With this result we can now construct capacity achieving PIR protocols for a wide range of system parameters.
\begin{theorem}[Capacity achieving PIR protocols] \label{th:subpacketization}
    For any given $\mu$-$(F,N)$ system with $\mu=t/N$,  $t\in [N]$ and file size $L=v\times t^F$, we can design a capacity achieving PIR scheme by using a $\bigg(\frac{N}{\gcd(N,t)},\frac{t}{\gcd(N,t)},N,t\bigg)$-configuration in Theorem~\ref{th:design2pir}.

\end{theorem}
\begin{IEEEproof}
Since the parameters $v=\frac{N}{\gcd(N,t)}$, $k=\frac{t}{\gcd(N,t)}$, $b=N$ and $r=t$ satisfy \eqref{existence config}, a configuration with the above parameters exists. 
The PIR scheme designed by using this configuration for the storage scheme in Theorem~\ref{th:design2pir}, and Sun-Jafar protocol for the retrieval of individual subfiles achieves capacity for the $\mu$-$(F,N)$ system with $L=v\times t^F$. This is true since the Sun-Jafar protocol achieves capacity for the $1$-$(F,t)$ systems with files size $t^F$. 
\end{IEEEproof}

\begin{remark}
The results shown in Theorem~\ref{th:design2pir} and Theorem~\ref{th:subpacketization} can be extended to arbitrary $\mu$, by using the concept of memory sharing as done in \cite{attia20}.
\end{remark}

\begin{remark}
Theorem~\ref{th:subpacketization} establishes that, we can achieve capacity for the $\mu$-$(F,N)$ system using a subpacketization $v\leq N$ i.e. the minimum subpacketization required by the proposed PIR scheme is linear in the number of servers $N$.
\end{remark}




\subsection{An Example}
Consider a $\frac{2}{3}$-$(2,3)$ system i.e., a system with 3 servers, $\mu={2}/{3}$ and 2 files $a$ and $b$, each 12 bits long.
We shall use the $(3,2,3,2)$-configuration with the incidence matrix $M$ given below
for the storage scheme. So, we divide each file into 3 subfiles ($a\to a_1,a_2,a_3$, $b \to b_1,b_2,b_3$), each 4 bits long.
Let {$a_{ij}$ and $b_{ij}$, ${\;i\in [3],j\in[4]}$} represent the $j^{th}$ bit of subfile $a_i$ and $b_i$ respectively. 
Store the subfiles on the servers as follows:
\begin{eqnarray}
M = \begin{bmatrix}1&1&0\\0&1&1\\1&0&1 \end{bmatrix} \quad 
\begin{array}{ccc}
 \hline
\Se_1 & \Se_2 & \Se_3 \\
\hline
a_1 & a_1 & a_2\\ 
b_1 & b_1 & b_2\\
a_3 & a_2 & a_3\\
b_3 & b_2 & b_3\\
\hline
\end{array}
\end{eqnarray}

This system can be projected onto the following reduced
$1$-$(2,2)$ systems.

\begin{table}[!htb]
    \begin{subtable}{.33\linewidth}
      \centering
        \caption{$\Se(W_{\theta 1})$}
            \begin{tabular}{cc}
            \hline
            $\Se_1$ & $\Se_2$ \\
            \hline
            $a_1$ & $a_1$ \\ 
            $b_1$ & $b_1$ \\
            \hline
            \end{tabular}   
    \end{subtable}%
    \begin{subtable}{.33\linewidth}
      \centering
        \caption{$\Se(W_{\theta 2})$}
            \begin{tabular}{cc}
            \hline
            $\Se_2$ & $\Se_3$ \\
            \hline
            $a_2$ & $a_2$ \\ 
            $b_2$ & $b_2$ \\
            \hline
            \end{tabular} 
    \end{subtable} 
    \begin{subtable}{.33\linewidth}
      \centering
        \caption{$\Se(W_{\theta 3})$}
            \begin{tabular}{cc}
            \hline
            $\Se_1$ & $\Se_3$ \\
            \hline
            $a_3$ & $a_3$ \\ 
            $b_3$ & $b_3$ \\
            \hline
            \end{tabular} 
    \end{subtable}     
\end{table}

Suppose the user wishes to retrieve file $a$. 
Subfile $a_1$ is retrieved from system $\Se(W_{\theta1})$ using the protocol in \cite{sun17} as follows: 
We start by querying server $\Se_1$ for bit $a_{11}$. Now to obfuscate $\Se_1$, we also demand bit $b_{11}$ from it. Similarly, we query server $\Se_2$ for bits $a_{12}$ and $b_{12}$. At this point we have 2 desired bits and 2 undesired bits. Now, we query the servers for linear combination of a unknown desired bit and known undesired bit. So, we query $\Se_1$ for $a_{13}+b_{12}$ and $\Se_2$ for $a_{14}+b_{11}$. Clearly, we can retrieve all the desired bits of subfile $a_1$ using these queries. Privacy can be maintained by using a random permutation of bits of $a_1$ and $b_1$ instead of using them in the original order. Subfiles $a_2$ and $a_3$ can be retrieved in a similar manner. The queries to be sent to servers $\Se_1$, $\Se_2$ and $\Se_3$ for retrieval of file $a$ are summarised as follows:

\begin{center}
\begin{tabular}{ccccc}
\hline
$\Se_1$ & &$\Se_2$ & &$\Se_3$ \\
\hline
$a_{11}$ & &$a_{12}$ && $a_{21}$\\ 
$b_{11}$ & &$b_{12}$ & &$b_{21}$\\
$a_{13}+b_{12}$& & $a_{14}+b_{11}$& & $a_{23}+b_{22}$\\ 
$a_{31}$ & &$a_{22}$ & &$a_{32}$\\
$b_{31}$ & &$b_{22}$ & &$b_{32}$\\
$a_{33}+b_{32}$& & $a_{24}+b_{21}$ & &$a_{34}+b_{31}$\\
\hline
\end{tabular}    
\end{center}

Suppose the queries for $\Se_2$ and $\Se_3$ were \{$a_{13}$, $b_{13}$, $a_{11}+b_{14}$, $a_{23}$, $b_{23}$, $a_{21}+b_{24}$ \}, \{ $a_{21}$, $b_{21}$, $a_{23}+b_{22}$, $a_{33}$, $b_{33}$, $a_{31}+b_{34}$ \} then file $b$ would be retrieved for the same set of queries for $\Se_1$. 
We can design similar queries for recovering file $c$. 
Therefore the queries to the servers do not leak information about the file requested.

Finally, note that the capacity for the given $\frac{2}{3}$-(2,3) system is $R^{*}(\mu)=\big(1+1/t \big)^{-1}=\big(1+1/2 \big)^{-1}={2}/{3}$.
The rate of the above PIR scheme is ${R}(\mu)= {12}/{18}={2}/{3}$, which matches the capacity of the given system.

\section{Performance of proposed protocol}\label{sec:performance}

 In this section we report the performance of the proposed PIR schemes.
 We also compare the performance of proposed protocol with previous work.

\begin{figure}
    \centering
    \includegraphics[scale=0.95]{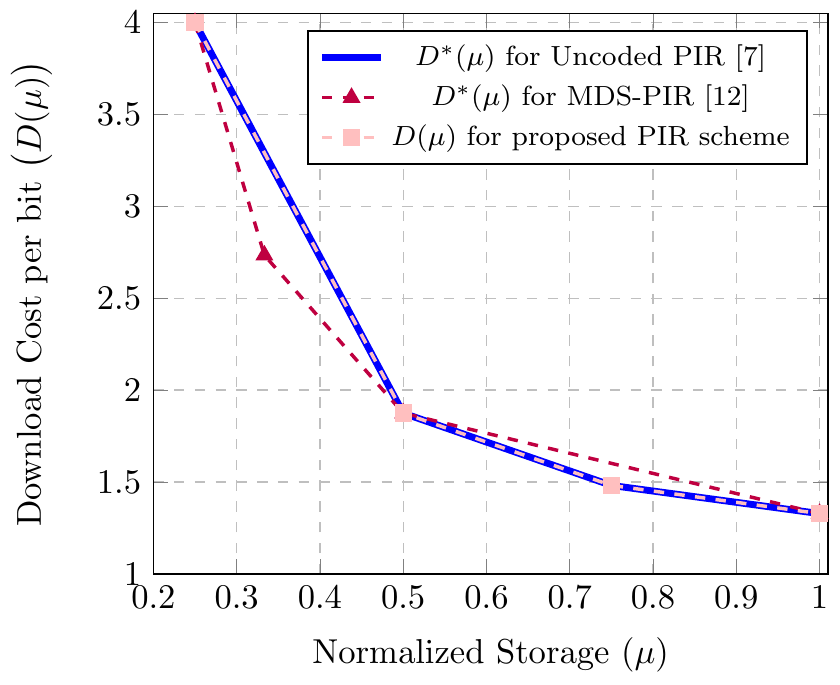}
    \caption{\small $D(\mu)$ \textit{vs.} $\mu$ for $\mu$-$(F=4, N=4)$ system. Also shown are the optimal download costs for MDS and uncoded PIR protocols.}
    \label{fig:2}
    \end{figure}

Fig.~\ref{fig:2} shows the variation of the download cost for the proposed protocol as a function of normalized storage $\mu$ for a $\mu$-$(F=4, N=4)$ system. 
It can be seen that the proposed protocol achieves capacity. 
In the same figure we also plot the optimal download cost (inverse of PIR capacity) for the given system using uncoded and coded PIR protocols, using \eqref{eq:pir-capacity} and
\eqref{eq:pir-cap-mds}.
%

\begin{figure}[H]
        \centering
        \includegraphics[scale=0.95]{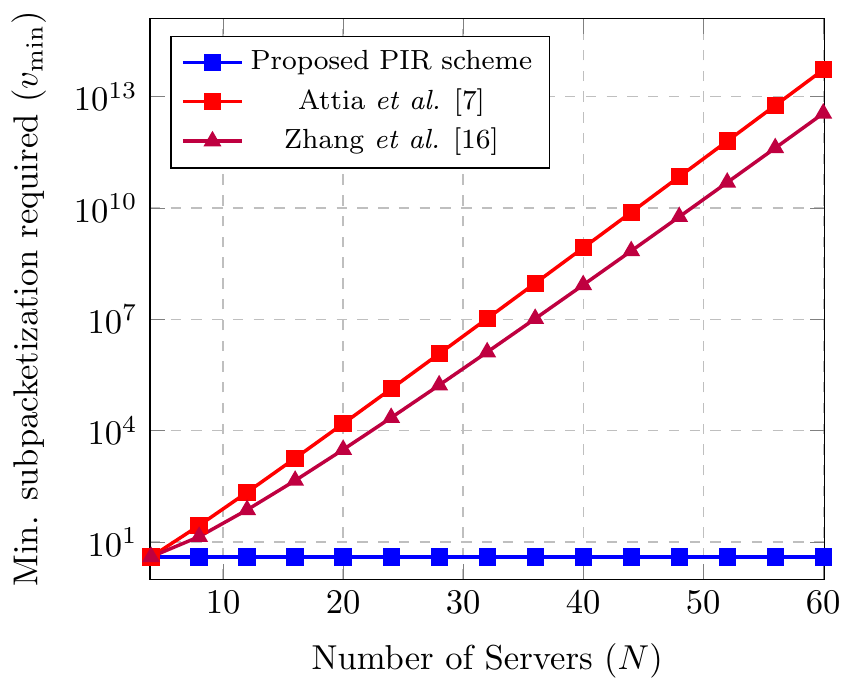}
        \caption{$v_{\min}$ \textit{vs.} $N$ for $\mu=1/4$}
        \label{fig:3}
    \end{figure}

Fig.~\ref{fig:3} shows a comparison of minimum subpacketization required to achieve capacity for these three schemes as a function of the number of servers $N$, keeping $\mu$ fixed.
The proposed protocol requires a minimum subpacketization of $v=\frac{N}{\gcd(N,t)}$, whereas the schemes in \cite{attia20} and \cite{zhang19} require minimum subpacketizations of $\binom{N}{t}$ and $\binom{N}{t}/{t}$ respectively. 
%

The minimum file size required to achieve capacity for the proposed protocol is given by $L=\frac{N}{\gcd(N,t)}t^F$. The minimum files size required to achieve capacity for the protocols proposed in \cite{attia20} and \cite{zhang19} are $L=\binom{N}{t}t^F$ and $L=\binom{N}{t}t^{F-1}$ respectively.
Table~\ref{tab:file-size} shows a comparison of the minimum file size required to achieve capacity for the three schemes. 
\begin{center}
\begin{table}
\renewcommand{\arraystretch}{1.2}
\caption{\small Minimum File size required to achieve capacity}
\label{tab:file-size}
\centering
\begin{tabular}{|c|c|c|c|c|c|c|}

\hline 
{$N$} & {$\mu$} & {$F$} &  \multicolumn{3}{c|}{Min. file size (MB)}
\tabularnewline
\cline{4-7} \cline{5-7} \cline{6-7} \cline{7-7} 
 &  &  & \cite{attia20}&  \cite{zhang19}& Proposed \tabularnewline
 &  &  & &  &  protocol\tabularnewline

\hline 
6 & 1/2 & 5  & 5.4 $\times 10^{-4}$& 1.8 $\times 10^{-4}$ & 5.4$\times 10^{-5}$
\tabularnewline
\hline 
7 & 3/7 & 5 &  $ 10^{-3}$ & $ 3.33 \times 10^{-4}$ & $2 \times 10^{-4}$ 
\tabularnewline
\hline 
18 & 1/2 & 8  & 2.43$\times 10^{5}$ & 2.70$\times 10^{4}$& 10.264 \tabularnewline
\hline 
190 & 1/10 & 5  & 1.85 $\times 10^{25}$ & 9.73 $\times 10^{23}$ & 2.45 \tabularnewline
\hline 
\end{tabular}
\end{table}
\end{center}

\section{Conclusion}\label{sec:conc}
In this paper we presented a novel capacity achieving PIR protocol for storage constrained uncoded PIR systems. Apart from achieving capacity, our protocol has additional advantage that it offers linear subpacketization in the number of servers as compared to the exponential subpacketization required in the previous works. Moving forward, one natural direction  is to establish bounds on the subpacketization required to achieve capacity. Developing efficient PIR protocols when file size is not a multiple of $v \times t^F$ is another interesting problem.  

\nix{
\nocite{*}
\bibliographystyle{ieeetr}
\bibliography{main_arxiv}}

\end{document}